\documentclass[5p, sort&compress]{elsarticle}

\pdfoutput=1

\usepackage{amsmath}
\usepackage{amssymb}
\usepackage{amsfonts}
\usepackage{hyperref}
\usepackage{color}

\usepackage{graphicx}

\begin{document}

\title{{\bfseries Anomalous Compliance and Early Yielding of Nanoporous Gold}}

\author[hzg,tud]{Bao-Nam~Dinh~Ng{\^o}\corref{cor}}
\ead{dinh.ngo@hzg.de}
\author[tud]{Alexander~Stukowski}
\author[hzg]{Nadiia~Mameka}
\author[hzg,tuhh]{J{\"u}rgen~Markmann}
\author[tud]{Karsten~Albe}
\author[hzg,tuhh]{J{\"o}rg~Weissm{\"u}ller}

\address[hzg]{Helmholtz-Zentrum Geesthacht, Institut f{\"u}r Werkstoffforschung, Werkstoffmechanik, Geesthacht, Germany}

\address[tud]{Technische Universit{\"a}t Darmstadt, Fachbereich Material- und Geowissenschaften, Fachgebiet Materialmodellierung, Darmstadt, Germany}

\address[tuhh]{Technische Universit{\"a}t Hamburg-Harburg, Institut f{\"u}r Werkstoffphysik und Werkstofftechnologie, Hamburg, Germany}

\cortext[cor]{Corresponding author. Tel: +49 4152 87 2611}

\begin{abstract}
We present a study of the elastic and plastic behavior of nanoporous gold in compression, focusing on molecular dynamics simulation and inspecting experimental data for verification. Both approaches agree on an anomalously high elastic compliance in the early stages of deformation, along with a quasi immediate onset of plastic yielding even at the smallest load. Already before the first loading,  the material undergoes spontaneous plastic deformation under the action of the capillary forces, requiring no external load. Plastic deformation under compressive load is accompanied by dislocation storage and dislocation interaction, along with strong strain hardening. Dislocation-starvation scenarios are not supported by our results. The stiffness increases during deformation, but never approaches the prediction by the relevant Gibson-Ashby scaling law. Microstructural disorder affects the plastic deformation behavior and surface excess elasticity might modify elastic response, yet we relate the anomalous compliance and the immediate yield onset to an atomistic origin: the large surface-induced prestress induces elastic shear that brings some regions in the material close to the shear instability of the generalized stacking fault energy curve. These regions are elastically highly compliant and plastically weak.
\end{abstract}
\maketitle
\section{Introduction}

Nanoporous gold (NPG) made by dealloying has attracted much attention due to its prospective applications in, e.g., actuation \cite{Kramer2004,Biener2009,Jin2010,Jin2010a,Detsi2011b,Detsi2012a}, catalysis \cite{Ding2004,Xu2007}, sensing \cite{Lavrik2001,Lang2009}, and microfluidics \cite{Xue2014}. The material represents a network of short fibers or `ligaments' with dimensions that can be controlled between 5 nm and several micrometers \cite{Li1992,Kertis2010,Wang2013}. Significantly, dealloying conserves the crystal lattice of the parent alloy (in the order of $10-100\:\mathrm{\mu m}$), so that ligaments are crystallographically coherent over large distances \cite{Parida2006, Jin2009}. Representing an ensemble of interconnected ligaments, NPG offers an implementation of the mechanical properties of nanowires, including specifically their high strength, into a macroscopic material. The material is thus a model system for study of the mechanical properties and the mechanisms of plastic deformation at the nanoscale.

Experiments, and specifically the tests of macroscopic samples to large compressive strain of Refs. \cite{Jin2009,Jin2011,Wang2013}, have established characteristic features of NPG deformation which are now briefly summarized and compared to recent simulation studies using molecular dynamics (MD) \cite{Sun2013,Farkas2013} and finite element modeling (FEM) \cite{Huber2014}.

NPG, while brittle in tension \cite{Li1992,Biener2005,Jin2009,Balk2009}, can be highly deformable in compression. Early experiments demonstrated this for focused-ion-beam (FIB) cut micropillars \cite{Volkert2006,Biener2006}, but improved synthesis protocols now allow deformation to large compressive strain for mm-sized samples \cite{Jin2009,Jin2011,Wang2013} that are free of artifacts from FIB cutting. Gibson--Ashby scaling equations \cite{Gibson1982,GibsonAshby1999} have been used for estimating the local yield stress of ligaments based on the effective materials behavior of NPG \cite{Biener2005a,Volkert2006}. However, the results of different measurements disagree. Micro and nanoscale tests suggest that the yield stress of ligaments increases with decreasing size, approaching the theoretical shear strength at sizes of a few nanometers \cite{Biener2005a,Volkert2006,Biener2006,Hodge2007,Hakamada2007,Dou2011}. The results for mm-sized samples confirm the trend, yet they reveal systematically lesser strength \cite{Jin2009,Balk2009}. This discrepancy partly reflects inconsistencies in converting nanoindentation hardness to yield stress \cite{Jin2009,Balk2009}. More importantly, however, the compression experiments testify to an extended elastic-plastic transition, with early yielding and with a strong initial increase of the flow stress \cite{Huber2014}. The yield strength is therefore poorly defined. MD studies have so far not reproduced this finding, indicating instead a pronounced linear elastic regime \cite{Sun2013} and compressive yield strength in agreement with the scaling equations \cite{Farkas2013}.

Most studies find NPG more compliant than predicted by the relevant Gibson--Ashby law \cite{BienerBook2007,Hodge2009,Huber2014}. There are, however, two reports of unusually high stiffness at small ligament size \cite{Mathur2007,Liu2013}. The recent tests on mm-sized samples show an unusually high initial compliance and a significant stiffening during the early stages of plastic deformation \cite{Huber2014,Mameka2014}. This evolution in elastic response cannot be explained by Gibson-Ashby type scaling of the effective stiffness with density \cite{Huber2014}. While microstructural disorder in the network of ligaments was found to favor high compliance \cite{Huber2014}, the mechanisms behind the unusually high initial compliance and the strong stiffening during early-stage deformation of NPG remain yet to be identified. In fact, large discrepancies between the reported experimental stiffness data call for an independent verification.

Experiments show practically no transverse strain during compressive plastic deformation under uniaxial load \cite{Jin2009}. This observation is reproduced by MD simulations and FEM \cite{Farkas2013,Huber2014}, with FEM linking the observation to microstructural disorder \cite{Huber2014}. As a consequence of the absence of transverse strain, uniaxial compression densifies the network. Consistent with this densification and with the scaling laws, compression brings an unusually strong strain hardening \cite{Jin2009}. It has been suggested that this implies a compression-tension asymmetry, with strain hardening in compression and strain softening in tension \cite{Wang2013}. The tensile brittleness of NPG might thus reflect the plastic instability that ensues from strain softening. Indeed, constraining the volume change by infiltrating the pores of NPG with a polymer, while strengthening the material,  suppresses the strain hardening and establishes ductility in tension \cite{Wang2013}. Tensile deformation of NPG has been probed by atomistic computer simulation, with one study confirming the tensile strain softening \cite{Sun2013} while another finds strain hardening \cite{Farkas2013}. The origin of this different behavior has not been addressed.

Dislocations are accumulated during plastic deformation of NPG and eventually even form cell walls. Electron backscatter diffraction detects the ensuing mosaic structure at length scales larger than the ligament size \cite{Jin2009}, while high-resolution transmission electron microscopy shows stacking faults and twins generated upon compression \cite{Dou2011}. Strain hardening, besides partly reflecting densification, is also affected by interactions between the stored dislocations. This is revealed by the comparison of experiment to results from finite element modeling \cite{Huber2014} and is also supported by the emergence of strain-rate sensitivity as the defect density increases during deformation \cite{Jin2009}.

In this study, we present a study of NPG compression deformation by MD simulations, addressing specifically the immediate plasticity and subsequent strain hardening, as well as the unexpectedly high initial compliance and subsequent stiffening. For verification, experimental data are shown alongside the simulations.

\section{Methods}
\subsection{Simulation}

A porous microstructure on a rigid crystal lattice was generated by spinodal decomposition of a binary mixture, using the Metropolis Monte Carlo algorithm with an Ising type Hamiltonian and periodic boundary conditions in all directions. A cubic box of $100 \times 100 \times 100$ lattice spacings ($408\:\mathrm{\AA}$) and $\{100\}$-oriented edges was randomly filled with atom types $A$ and $B$ at $30\:\%$ atomic fraction $A$. Interaction energy for nearest neighbor atoms was zero for pairs of the same type, and $0.02\:\mathrm{eV}$ for pairs of different types. After the phase separation, all $B$-type atoms were removed. All isolated $A$-type atoms were also removed to assure a connected structure. As a result, the solid (volume- and mass-) fraction, $\varphi$, was $0.297$. The same procedure was applied to $B$-type atoms to ensure open porosity.

For the molecular dynamics simulations we used the open-source code \textsc{LAMMPS} \cite{Plimpton1995} with an EAM potential for gold \cite{Foiles1986}. In most simulations the temperature was maintained at $300\,\mathrm{K}$, while selected runs explored low temperature conditions ($T=0.01\:\mathrm{K}$) to assess the conceivable role of thermally activated processes.

In each run, the energy of initial structure was first minimized by the conjugate gradient method. To ensure a force-free structure after energy relaxation, the minimization was performed until the specified force tolerance is less than $10^{-4}\:\mathrm{eV/\AA}$, while the energy criterion was set to zero. At the point where the minimization converged, the relative change in energy was less than $10^{-12}$. Then, the structure was thermally relaxed for $1\:\mathrm{ns}$ at zero load. Uni-axial compression was attained by scaling the whole simulation box in one dimension at each time step at the engineering strain rate of $10^8\:\mathrm{s}^{-1}$, while keeping other dimensions at zero stress. Temperature and pressure were maintained by a Nos\'{e}-Hoover thermostat and barostat \cite{Nose1984,Hoover1985} with Martyna-Klein-Tuckerman modifications \cite{Martyna1992,Martyna1996}. The time step for all simulations was $2\:\mathrm{fs}$. All strains in this work are specified as  compressive engineering strain, $\varepsilon$.

The evolution of Young's modulus of NPG was obtained by carrying out another simulation, using similar conditions but with intermediate unload/reload cycles. The effective Young's modulus was determined as a tangent modulus \cite{astm-e111} from the unload stress-strain segments.

The open-source software \textsc{Ovito} \cite{Stukowski2010} was used for visualization. Dislocation defects in the material were detected and identified with the dislocation extraction algorithm (DXA) \cite{Stukowski2010a,Stukowski2012}. This computational tool allows us to measure dislocation densities (broken down by Burgers vector) during the simulation. A surface reconstruction algorithm \cite{Stukowski2013} was used to generate a geometric representation of the evolving crystal surfaces from the atomic positions and measure surface area and solid volume. Planar defects such as  stacking faults and twin boundaries were detected and classified by means of an atomic pattern matching algorithm \cite{Stukowski2013} that is based on the adaptive common neighbor analysis (a-CNA) method \cite{Stukowski2012a}.

\subsection{Experiment}

NPG specimens for the experimental study were synthesized by electrochemical dealloying as described in Ref.~\cite{Jin2011}. The processing of the $\mathrm{Au}_{25}\mathrm{Ag}_{75}$ alloy precursor for the dealloying included arc melting, homogenization at $900\,^{\circ}\mathrm{C}$ for 120~h in vacuum, wire drawing, and cutting to form cylindrical samples of $1.19-1.21$~mm in diameter and $1.90-2.00$~mm in length. These samples were then annealed in inert atmosphere ($800\,^{\circ}\mathrm{C}$, 2~h) and used for dissolution at 0.75~V in 1~M $\mathrm{HClO}_4$ electrolyte at room temperature. Following this, a polarization at 1.1~V until the current decayed to 10~$\mu \mathrm{A}$ and potential cycling at a scan rate of 5~mV/s between $-0.5$~V and $1.0$~V for 20 cycles was performed on as-dealloyed NPG in fresh 1~M $\mathrm{HClO}_4$ solution. This served to remove residual Ag as well as strongly bound oxygen from the surface. Finally, the NPG samples were immersed in deionized ultra-pure water for at least 12~h and left for drying in Ar flow for more than two days. Scanning electron microscopy (SEM) yielded a ligament size of $40\pm 5$~nm in the as-prepared specimens. The initial solid fraction was estimated at $\varphi = 0.298\pm 0.01$, based on the sample dimensions and mass. Similar to previous study \cite{Jin2009}, the experimental sample is polycrystalline with no apparent crystallographic texture. With grain sizes in the order of $50\,\mu$m, the individual crystallites are three orders of magnitude larger than the ligaments.

Compression testing used a mechanical testing rig (Zwick Z010 TN) at constant engineering strain rate $10^{-4}\:\mathrm{s}^{-1}$ and under ambient conditions. A stress-strain curve was accompanied by unload/reload cycles to obtain the effective stiffness. The minimal stress for each unloading step was $1.75$~MPa. A detailed description of the stiffness evaluation can be found in Ref.~\cite{Huber2014}.

The removal of oxygen species from the surface of NPG concurs with coarsening. As adsorbed oxygen has a strong influence on the mechanical behavior of NPG \cite{Jin2011,Mameka2014}, we here compare mechanical response of the virtual sample with that of samples which, albeit having much bigger ligament size, exhibit clean surfaces.

\section{Results}
\subsection{Morphology and initial relaxation}

\begin{figure}[h!]
\centering
\includegraphics[scale=1]{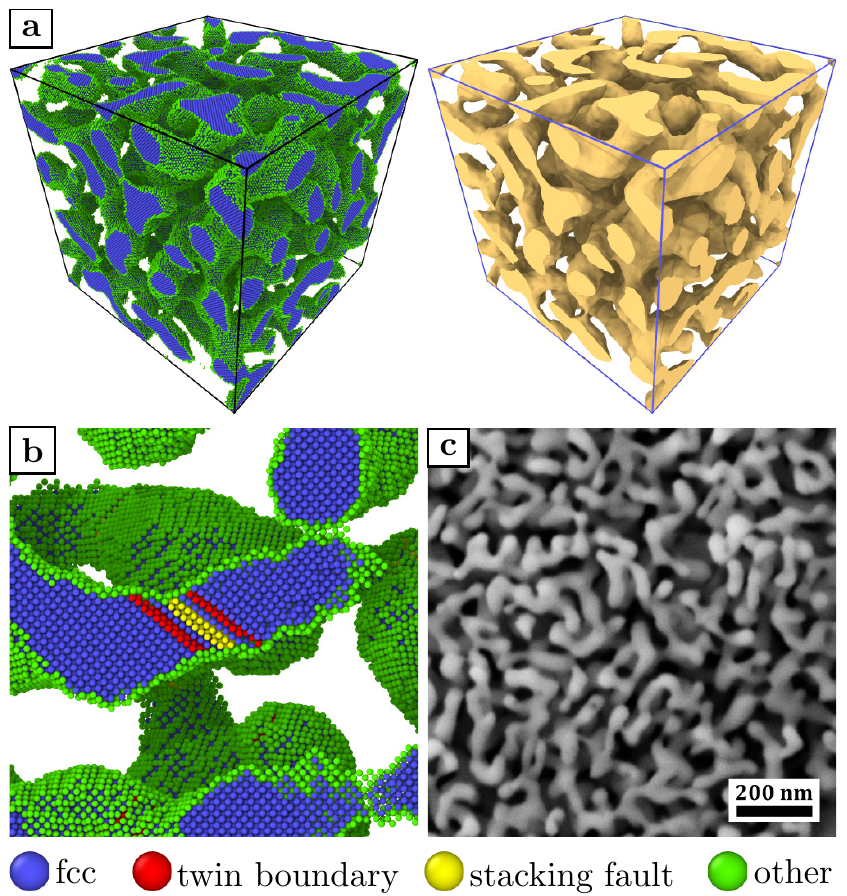}
\caption{
\label{fig:1}(Color online)
Nanoporous gold (NPG) structures used in this study. (a) As-prepared structure from spinodal decomposition (left) and its surface reconstruction (right). This sample has a relative density of $0.297$ and ligament size of $3.15\:\mathrm{nm}$. Capillary forces induce plasticity, letting the sample volume decreased by $3.6\:\%$ after equilibration. Thus, lattice defects can be observed in the relaxed structure before deformation. Some examples of those defects are shown in (b). Throughout this work, if not otherwise stated, fcc, twin boundary, stacking fault, and unclassified atoms are coded in blue, red, yellow, and green, respectively (see legend). (c) Scanning electron micrograph of the microstructure of a NPG sample. Relative density is $0.298\pm 0.01$, ligament size is $40\pm 5$~nm. Note the similarity to the microstructure of the virtual sample.}
\end{figure}

Figure \ref{fig:1}a shows a snapshot of the virtual sample  as generated  by the Monte Carlo simulation (left), as well as its surface reconstruction (right). Similar to previous MD studies, where a phase-field spinodal decomposition approach was used for generating the boundaries of the ligaments \cite{Farkas2013,Sun2013}, we find that our sample reproduces the interconnected nanoscale network structure of nanoporous gold well (see the SEM image in Fig. \ref{fig:1}c).

The specific surface area (area per volume of solid) of our virtual sample is $ \alpha =  1.05/\mathrm{nm}$. In order to relate $\alpha$ and $\varphi$ to a characteristic structural length-scale, $L$, we used the conversion rule:
\begin{equation}
\label{eqn:L_conversion}
L = \frac{1.63(1.25-\varphi)(1.89+\varphi(0.505+\varphi))}{\alpha},
\end{equation}
which yields $L = 3.15\:\mathrm{nm}$. This rule results from modeling NPG as a periodic diamond-like structure with cylindrical ligaments of diameter $L$, connected in spherical nodes which are just large enough to just cover the overlapping cylinder segments \cite{Huber2014}. For the experiment, electrochemical capacity measurements \cite{Cattarin2009} yield the volume-specific surface area of the tested sample, $\alpha = 0.0923\pm 0.002/\mathrm{nm}$. The conversion rule given in Eq.~(\ref{eqn:L_conversion}) thus puts the ligament diameter of the respective diamond-like structure at $L=35.8\pm 0.8\,\mathrm{nm}$, which is well compatible with the value estimated from the SEM image.

At $\sim 40$~nm edge length, our simulation box dimensions are one order of magnitude larger than the ligament size of $\sim3$~nm. This suggests that the box size is sufficient to rule out finite-size effects in the MD simulation. For verifying that the simulation box contains a statistically representative volume of the virtual nanoporous structure generated by spinodal decomposition, we created several virtual samples of same density and structure size by independent runs of spinodal decomposition under identical conditions. All virtual samples were found to exhibit consistent mechanical behavior, supporting the adequate size of the box.

\begin{figure}[h!]
\centering
\includegraphics[scale=1]{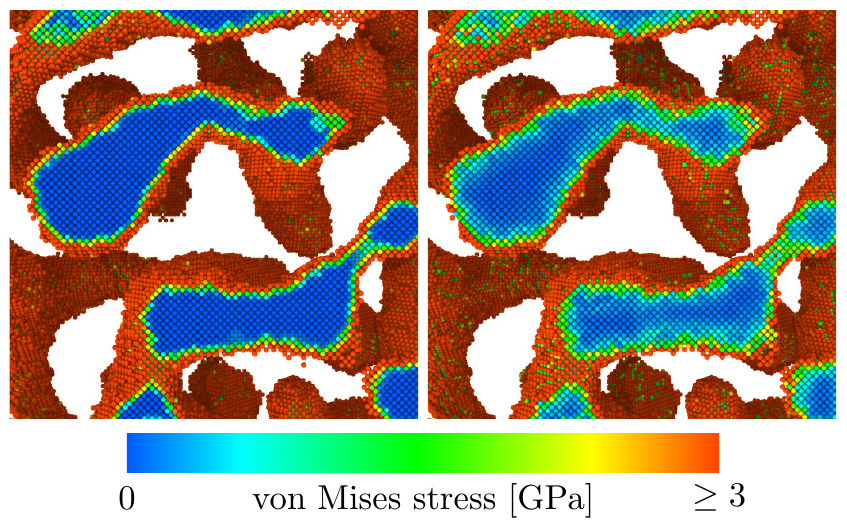}
\caption{
\label{fig:2}(Color online) Surface stress influence on local stress distribution. This figure shows a slice cutting through NPG sample before (left) and after (right) the initial energy minimization, with atoms color-coded according to the local von Mises stress. Note color change in the bulk regions upon energy relaxation, indicating local shear stress balancing the surface stress.}
\end{figure}

Relaxing the initially rigid and defect-free porous crystal at $T=0.01\,\mathrm{K}$ simulation and with no external load led to $2.96\:\%$ decrease in the total volume, increasing the effective solid fraction to $\varphi=0.306$. Not a single dislocation could be found in the relaxed structure, but  the relaxation generated a very small number of stacking faults and twin boundaries. The fault density was $\sim 7 \times 10^{-5} \rm nm^{-1}$, corresponding to only $\sim 120$ atoms in faulted configurations, a negligible fraction of the $\sim 1.2$ million atoms in the entire structure. This suggests that the volume contraction after relaxation at low temperature is almost entirely elastic and can be understood as a natural consequence of the action of the surface stress. While the average surface-induced prestress can be calculated by using generalized capillarity equation (Section 4.1 below), the influence of surface stress is readily observable by monitoring the stress distribution. As an example, Fig.~\ref{fig:2} shows a slice cutting through NPG sample before (left) and after (right) the initial energy minimization, with atoms color-coded according to the local von Mises stress (assuming equilibrium atom volume for all atoms). It is evident that although lattice defects have not shown up yet, the bulk atoms already experience changes in local stress, with accentuating changes near the surface or at the smaller segments of ligaments, implying that, before the onset of straining, the relaxed structure already suffers large stress (with non-uniform distribution) induced by the surface stress. We note that Fig.~\ref{fig:2} agrees with previous MD study of NPG \cite{Farkas2013} where relaxation of initially unstrained NPG samples leads to changes in the corresponding histogram of local stresses.

As compared to the relaxation at $T=0.01\,\mathrm{K}$, equilibrating the starting structure at $300$~K brings more pronounced structural changes. Figure~\ref{fig:1}b illustrates that many lattice defects are generated; the densities of dislocations, stacking faults and twins are $4.8 \times 10^{-4} \mathrm{nm}^{-2}$, $3.2 \times 10^{-3}/ \mathrm{nm}^{-1}$ and $1.4 \times 10^{-3} \mathrm{nm}^{-1}$, respectively. The creation of lattice defects  even in the absence of an external load  is also observed in experiments. Transmission electron microscopy (TEM) reveals this process for NPG when ligaments yield to the surface-induced stress during fast dealloying \cite{Parida2006}.

The 300 K equilibration increases the solid fraction to $\varphi = 0.308$. The corresponding specific surface area and ligament diameter are $\alpha = 1.04/\mathrm{nm}$ and $L = 3.16\:\mathrm{nm}$, respectively. The equilibration creates a stable structure, and in particular - apart form the generation of lattice defects - the initial long-range coherent crystal lattice is maintained. Bearing in mind the periodic boundary conditions, one can consider the virtual NPG structure as an infinitely extended single crystal with pores and dislocations.

\subsection{Stress-strain behavior under compression}

\begin{figure}[h!]
\centering
\includegraphics[scale=1]{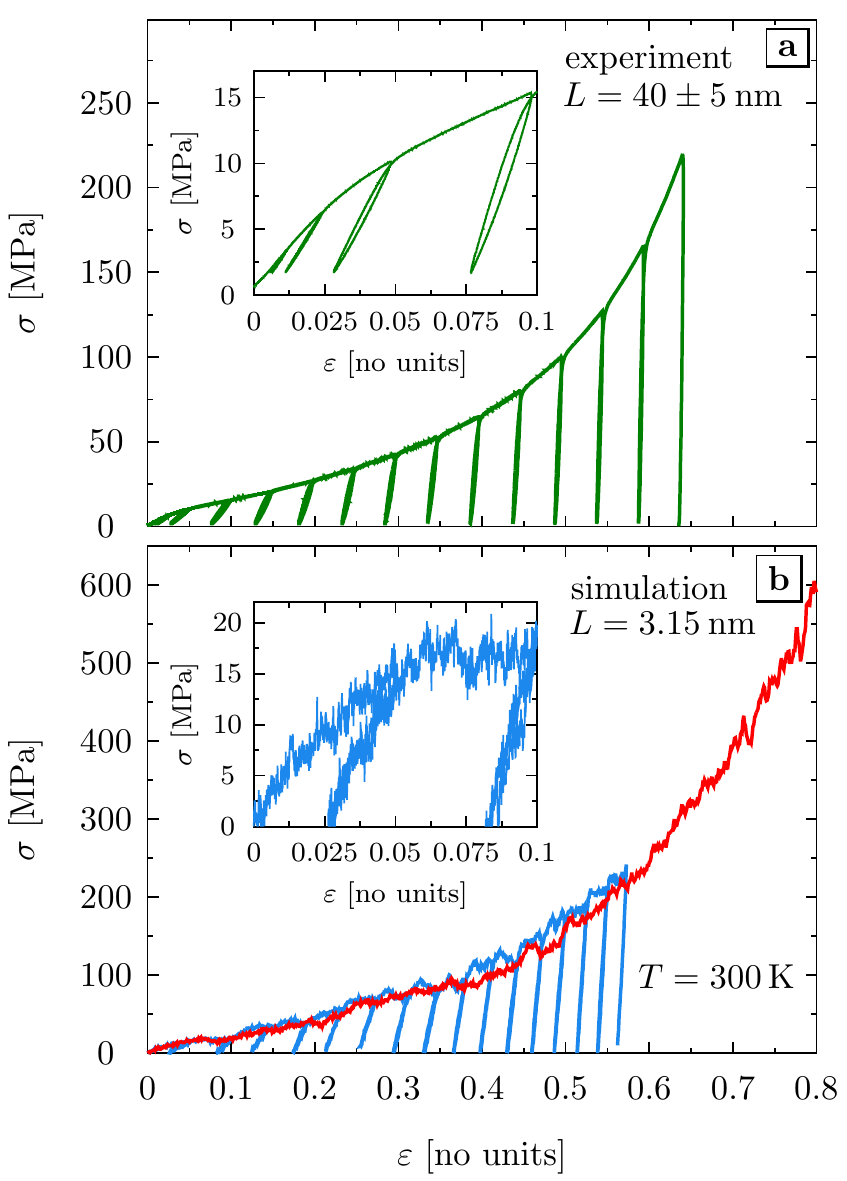}
\caption{\label{fig:3}(Color online) Deformation curve of nanoporous gold. (a) Experimental compressive stress-strain curve of a millimeter-sized sample subjected to a sequence of loading-unloading. (b) Simulated stress-strain curves of virtual sample at $300$~K. While the red curve depicts behavior of this sample in continuous compression to the axial engineering strain of $80\:\%$, the blue one presents its corresponding behavior in a sequence of load/unload. The inset magnifies the load-unload curve at strains less than $10\:\%$. Throughout this work, $\sigma$ and $\varepsilon$  denote compressive stress and axial engineering strain, respectively.}
\end{figure}

Next we studied stress-strain behavior in compression. As a background for inspecting the simulation results, Fig.~\ref{fig:3}a shows the experimental stress-strain curve for a mm-sized NPG sample with $L = 40\pm 5\,\mathrm{nm}$ described above. The simulated stress-strain curves of the virtual sample at 300~K are shown in Fig.~\ref{fig:3}b, with the continuous loading in red and the load/unload sequence in blue.  For the purpose of clarity, data at strains $ \leq 0.1$ are magnified in the insets.

The experimental data in Fig.~\ref{fig:3}a are well compatible with data for a sample of similar $L$ reported earlier in Ref.~\cite{Jin2009} (cf. Fig.~1d there). Remarkably, despite the deviation in ligament size between experiment and simulation, the respective stress-strain curves show excellent agreement. The small-strain deformation is nonlinear with a quasi immediate onset of plastic deformation already at the smallest load. This immediate yielding is suggested by the residual strain which is found upon unloading even after small deformation.

Experiment and simulation also agree in relation to the unusually pronounced strain hardening at larger strain. For example of the virtual sample, the flow stress under continuous loading is as low as $18.2\:\mathrm{MPa}$ at strain $\varepsilon=0.1$, rising to $93.1\:\mathrm{MPa}$ at strain $\varepsilon=0.4$ and finally attaining $591\:\mathrm{MPa}$ at the end of compression.

\begin{figure}[h!]
\centering
\includegraphics[scale=1]{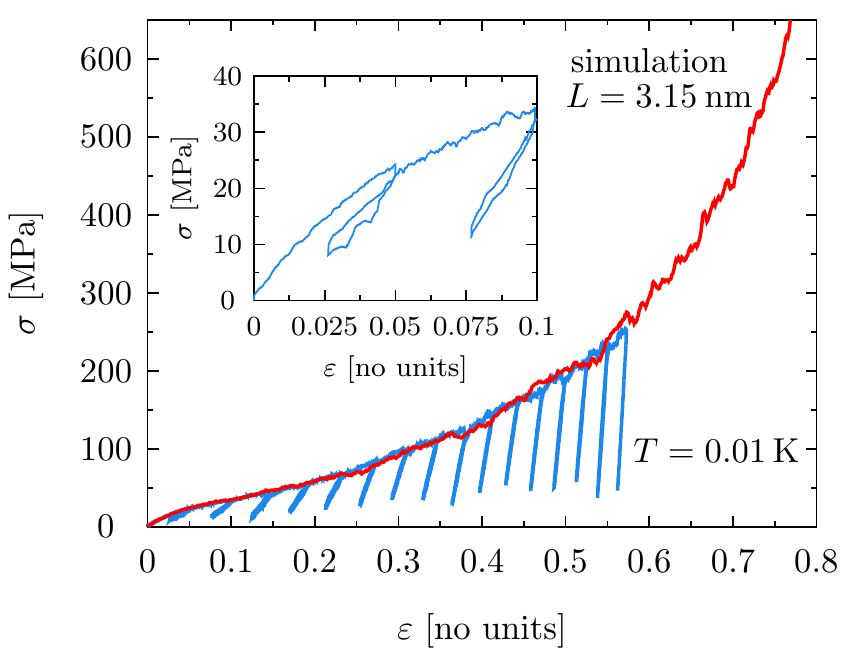}
\caption{\label{fig:4}(Color online) Deformation curve of nanoporous gold. These simulated stress-strain curves were recorded at $T=0.01\,\mathrm{K}$. As in Fig.~\ref{fig:3}b, the red and the blue curves represent the behavior under continuous loading and load/unload sequence, respectively. While the surrounding temperature differs significantly (0.01\,K vs. 300\,K), similar behavior is obtained, as is clearly seen here and in Fig.~\ref{fig:3}b}
\end{figure}

The  stress-strain curves of the virtual sample at $T=0.01\,\mathrm{K}$ are shown in Fig.~\ref{fig:4}, with graphs color-coded as in Fig.~\ref{fig:3}b. The stress-strain curves of virtual NPG are found for all practical purposes independent of the temperature.

We emphasize the deformability of NPG, real and virtual, in compression, along with low yield strength and high strain hardening. To elucidate the underlying mechanisms, we now inspect the deformation-induced structural changes, concentrating on the deformation in continuous compression.%

\subsection{Transverse strain and surface area}

\begin{figure}[h!]
\centering
\includegraphics[scale=1]{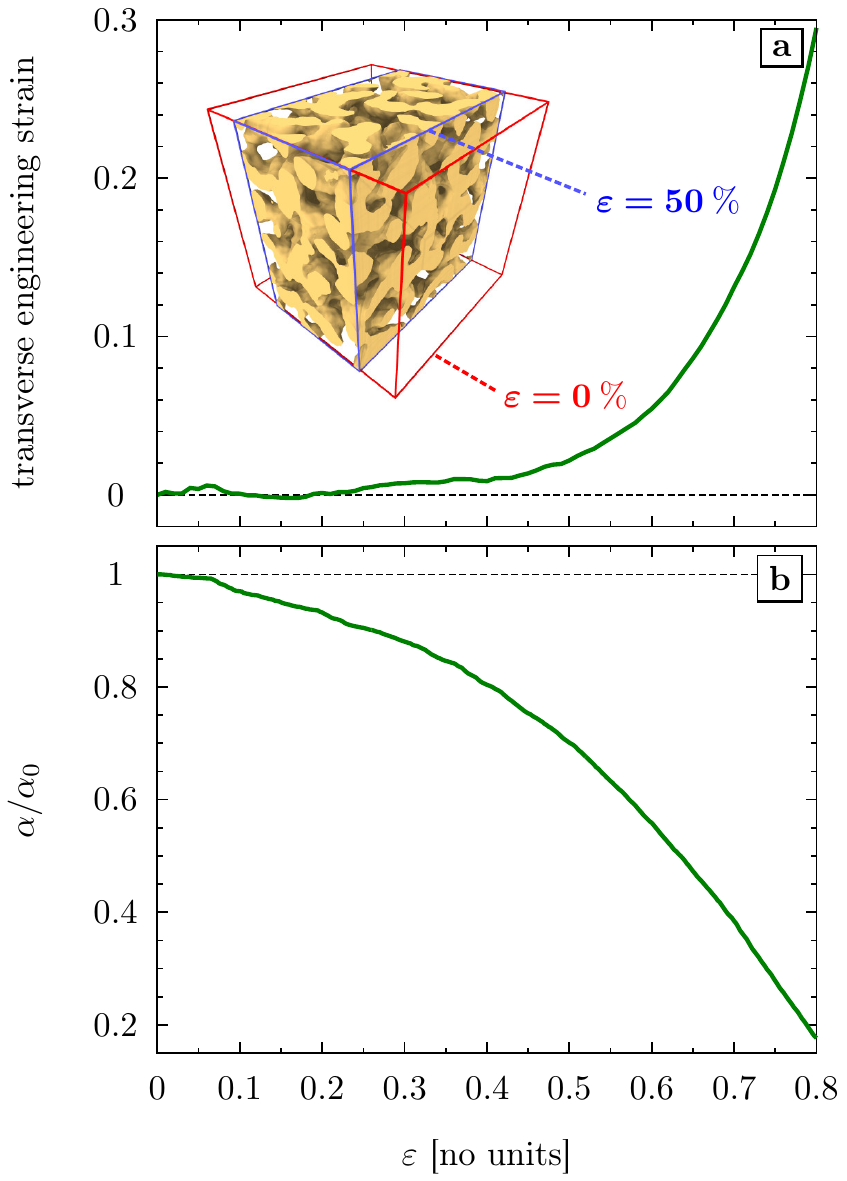}
\caption{\label{fig:5}(Color online) Structural change of NPG in compression. (a) Change of transverse strain with respect to axial strain of nanoporous gold in compression. Lateral expansion is negligible for strain up to $50\:\%$, as illustrated in the inset. (b) Change of surface area per solid volume, $\alpha$, of nanoporous gold in compression. A continuous loss of surface area is observed, as shown in Fig.~\ref{fig:5}b. All values are scaled to the initial value of the undeformed structure, $\alpha _0$. Entire evolution of reconstructed surface during deformation is shown in Movie-1 in the  Supplementary data.}
\end{figure}

Figure~\ref{fig:5}a shows the evolution of transverse strain during compression in our uniaxial load simulation. It is seen that the lateral dimensions remain essentially invariant for $\varepsilon$ up to $0.5$, which is in agreement with experimental data measured during plastic flow \cite{Jin2009,Wang2013}. Consequently, the change in axial dimension translates directly into volume reduction. In other words, compression entails densification. By the end of straining (at $\varepsilon = 0.8$) the solid fraction of the virtual sample attains the value of $\varphi = 0.918$.

One of the signatures of microstructural evolution is the variation of the specific surface area. The graph of $\alpha$ - scaled to the initial value, $\alpha _0$, of the structure before straining - is shown in Fig.~\ref{fig:5}b (entire evolution of reconstructed surface during deformation is shown in Movie-1 in Supplementary data). The data set reveals a continuous loss of surface area during compression.

\subsection{Evolution of effective Young's modulus}

\begin{figure}[h!]
\centering
\includegraphics[scale=1]{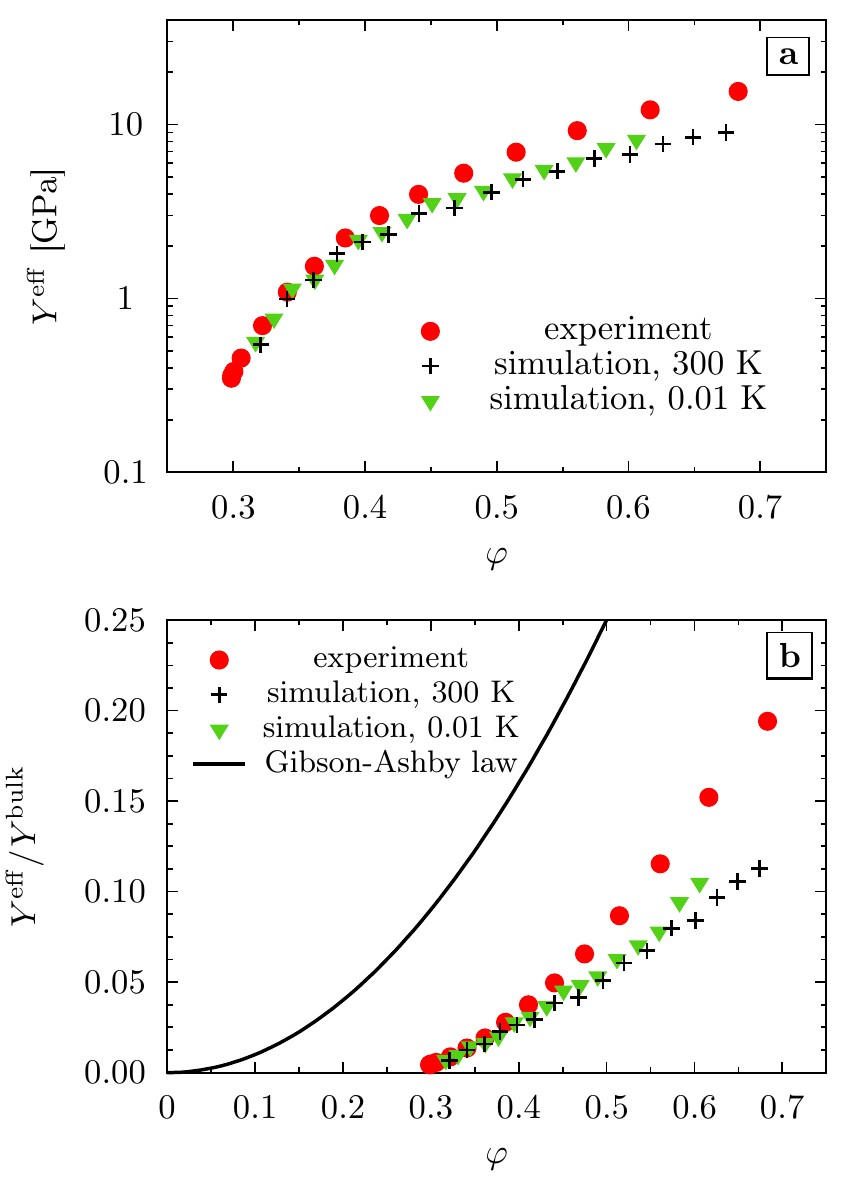}
\caption{\label{fig:6}(Color online) Change of effective Young's modulus, $Y^{\mathrm{eff}}$, of NPG during compression. While Fig.~\ref{fig:6}a shows the absolute values of $Y^{\mathrm{eff}}$ versus relative density, $\varphi$, in a log-linear plot, scaled values of $Y^{\mathrm{eff}}$ (to the Young's modulus $Y^{\mathrm{bulk}}$ of massive polycrystalline gold) are shown versus $\varphi$ in a linear-linear plot (Fig.~\ref{fig:6}b). Gibson-Ashby scaling law is also superimposed in Fig.~\ref{fig:6}b. The agreement between simulation and experiment is excellent. As is clearly seen in this plot, the evolution of $Y^{\mathrm{eff}}$ is characterized by two features: exceptionally high initial compliance and ensuing stiffening under compression. At any time during compression, NPG is much more compliant than the prediction of Gibson-Ashby law.}
\end{figure}

The load/unload segments of the experimental and simulated test yield data for the effective Young's modulus, $Y^{\mathrm{eff}}$, which we computed as an unloading tangent modulus \cite{astm-e111}. We verified that secant moduli, which average over the unload/reload process, provide consistent results. As the virtual samples densify during compression, we found it of interest to plot $Y^{\mathrm{eff}}$ versus the solid fraction. Figure~\ref{fig:6}a shows $Y^{\mathrm{eff}}$ and $\varphi$ in log-linear plots, revealing that the virtual sample starts out exceedingly compliant at both temperatures of the MD simulations. The stiffness then increases substantially as the compression proceeds. For example, at room temperature, while the first value of $Y^{\mathrm{eff}}$ at $5\:\%$ strain is $542\,\mathrm{MPa}$, the last value of $Y^{\mathrm{eff}}$ at $57\:\%$ strain is $9.03\:\mathrm{GPa}$, nearly 20-fold stiffer.

The experimental effective stiffness is shown along with the simulation data for comparison (see Fig.~\ref{fig:6}, circles). In view of the unusual compliance found by MD we emphasize the remarkable precision with which the simulation matches the experimental data, particularly at low strain where the compliance is highest.

The relation between effective modulus and relative density of macroscopic metal foams is the subject of one of the Gibson-Ashby scaling laws \cite{Gibson1982}:
\begin{equation}
\label{eqn:Gibson_Ashby}
Y^{\mathrm{eff}}=Y^{\mathrm{bulk}}\varphi^2 \,,
\end{equation}
where $Y^{\mathrm{bulk}}$ denotes Young's modulus of the bulk metal. We emphasize that this equation describes the variation in stiffness with the {\it initial} density of the foam structure. It is well established that the stiffness variation with density {\it during compression} follows quite different trends, see for example Ref. \cite{McCullough1999}. Nonetheless, it is of interest to compare our data to Eq.~(\ref{eqn:Gibson_Ashby}) as a reference for the behavior of porous metals. As a basis we calculated the Young's modulus of nontextured massive polycrystalline Au for the interatomic potential of our simulation at $0\:\mathrm{K}$ via Kr{\"o}ner's formulation \cite{Kroener1958}. The result, $Y^{\mathrm{bulk}}=78\:\mathrm{GPa}$, agrees well with the experimental value of massive polycrystalline Au at room temperature, $Y^{\mathrm{bulk}}=80\:\mathrm{GPa}$ \cite{Davis1998}. Therefore, we take $Y^{\mathrm{bulk}}=78\:\mathrm{GPa}$ and $Y^{\mathrm{bulk}}=80\:\mathrm{GPa}$ as the references for $Y^{\mathrm{bulk}}$ at $T=0.01\,\mathrm{K}$ and room temperature, respectively.

As an illustration for the first value of $Y^{\mathrm{eff}}$ at $300\:\mathrm{K}$, we substitute the corresponding solid fraction, $\varphi = 0.321$, into Eq.~(\ref{eqn:Gibson_Ashby}). The result is $8.2\:\mathrm{GPa}$ for $Y^{\mathrm{eff}}$, fifteen times higher than the value indicated by our simulation at that density. Consistently, at the first experimental value of the density,  $\varphi=0.299$, the scaling law overestimates  the experimental stiffness, for which we find $347\:\mathrm{MPa}$, by more than the factor twenty.

Figure~\ref{fig:6}b shows the evolution of $Y^{\mathrm{eff}}$ normalized to $Y^{\mathrm{bulk}}$, including the graph of Eq.~\ref{eqn:Gibson_Ashby} superimposed to the simulation results. The figure illustrates that the effective stiffness of NPG is consistently very much lower than that of the scaling law. It is not unusual for metal foams to be more compliant  than the prediction by the scaling law,  see for instance Ref. \cite{Michailidis2011b}, which reports a deviation by the factor three, and references therein. Yet, the magnitude of the deviation, which is apparent in our data (Fig.~\ref{fig:6}b) is extraordinarily high.

In relation to the trends in Fig.~\ref{fig:6}b we also find it remarkable that the stiffness values of our real and virtual samples appear to extrapolate to zero at a finite solid fraction, slightly less than 0.3. This trend is clearly distinguished from the power-law scaling  embodied in Gibson-Ashby type laws, and it illustrates the extraordinary compliance of NPG.

\subsection{Evolution of defect density}

\begin{figure}[h!]
\centering
\includegraphics[scale=1]{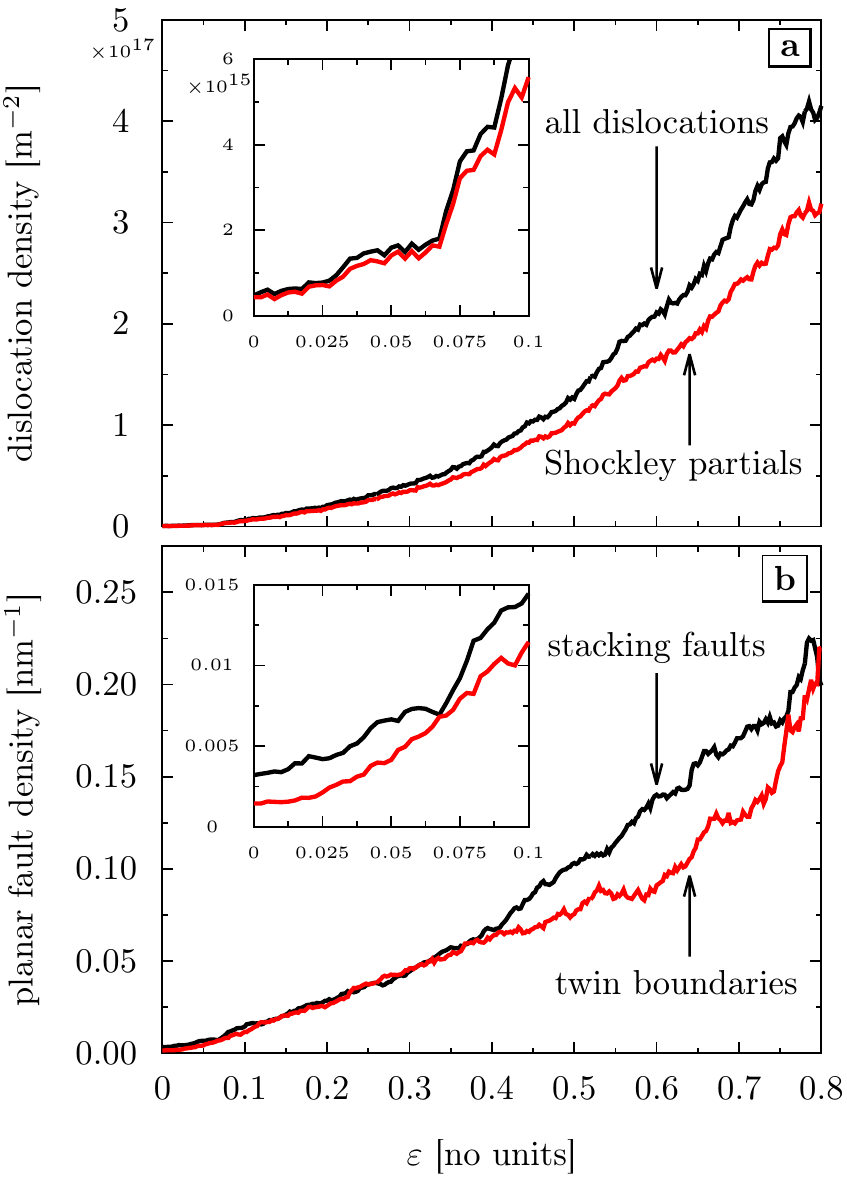}
\caption{\label{fig:7} (Color online) Evolution of fault density. (a) Accumulation of dislocation density during compression at $300\,\mathrm{K}$. Most contribution is from Shockley partials. (b) As a result of dislocation activity, twin boundaries and stacking faults populate during compression. The insets magnify parts of curves at strain less than $10\,\%$. As is clearly seen from these insets, dislocations are already present before the onset of straining, and get their activity immediately upon loading.}
\end{figure}

Figure~\ref{fig:7}a shows the evolution of dislocation density during compression. As is illustrated in the inset, Shockley partial dislocations -- generated during relaxation as stated earlier in this work -- are present already before the onset of loading. Compression then brings a continuous accumulation of dislocations. As the deformation proceeds the fraction of full dislocations increases, yet the Shockley partials remain predominant throughout at all strains.

\begin{figure}[h!]
\centering
\includegraphics[scale=1]{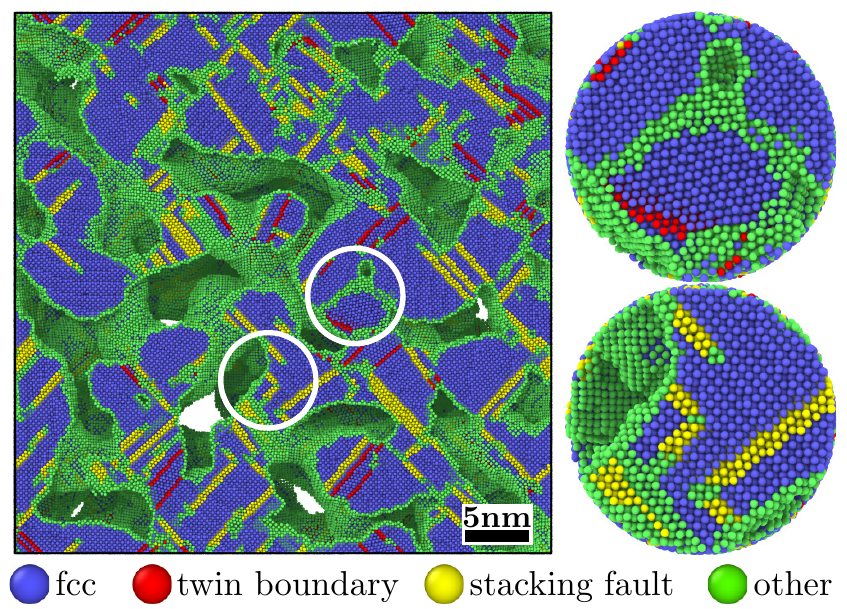}
\caption{\label{fig:8}(Color online) Typical defect structures found in nanoporous gold under compression. Left plot shows a slice cutting through sample at $55\:\%$ strain. Loading direction is perpendicular to plot plane. Magnifications of circled regions are shown on the right, depicting examples of grain boundaries (top) and Lomer-Cottrell locks (bottom) found in the deformed structure. Plasticity happens not only at the ligament junctions, but also in the middle of ligaments. Color coding is shown directly within the figure.  Entire evolution of defect structures during deformation is illustrated in Movie-2 and Movie-3 in the  Supplementary data.}
\end{figure}

The largest dislocation density, at the end of the deformation, is $\sim 4 \times 10^{17}\:\mathrm{m}^{-2}$ or $\sim 0.4\:\mathrm{nm}^{-2}$. Taking the characteristic volume of ligaments in the order of $L^3$, we infer few (in the order of 4) dislocations per ligament. This number is well compatible with the snapshot of defect configuration on a cross-section through a virtual sample at $55 \%$ engineering strain, Fig.~\ref{fig:8}.

The inset in Fig.~\ref{fig:7}a shows that the dislocation density increases immediately as the deformation sets in, even in the initial stages of compression. This observation represents another manifestation of the immediate yielding, confirming the conjectures obtained from the absence of linear elastic stress-strain relation and from the lasting plastic deformation after unloading from few percent net elastic-plastic deformation.

Figure~\ref{fig:7}b is analogous to Fig.~\ref{fig:7}a, except that the densities of twin boundaries and stacking faults are plotted. The analogous observations apply, with both densities increasing right from the start of loading and throughout the deformation.

A typical defect structure under compression is shown in Fig.~\ref{fig:8} (entire evolution of defect structures during deformation is illustrated in Movie-2 and Movie-3 in Supplementary data). In line with previous MD studies \cite{Sun2013,Farkas2013}, many twin boundaries (red) and stacking faults (yellow) are observed. Lomer-Cottrell locks are also found, as illustrated in the lower magnification in Fig.~\ref{fig:8}.

Figure~\ref{fig:8} also reveals that, in contrast to some experimental observations \cite{Sun2007,Dou2011}, planar defects were found not only in the ``nodes'' which connect the ligaments, but also in the middle of the ligaments.

At larger strains, grain boundaries form during densification, as exemplified by the upper magnification in Fig.~\ref{fig:8}. These boundaries result from cold-welding of neighboring ligaments. While this process contributes to densifying the structure, it is remarkable that the pore network, and with it the nanoporous structure, remains well-defined up to very large compressive strain, in spite of the largely increased solid fraction.

\section{Discussion}

Let us start out the discussion of the stress-strain behavior of NPG by a comparison to the behavior of macroscopic metal foams. Experiment, for instance Ref.s \cite{McCullough1999,Michailidis2011a,Michailidis2011b}, on these materials suggests stress-strain curves in compression quite similar to NPG. Specifically, foams with higher initial density show the same type of progressive hardening with ongoing deformation as NPG. However, closer inspection reveals significant differences. Firstly, conventional metal foams deform elastically at small stress, contrary to NPG. Secondly, load-unload tests on conventional foams show the effective stiffness to remain roughly constant during compression or to even slightly decrease in its initial stages \cite{McCullough1999}. This contrasts with the pronounced stiffening of NPG upon compression that is observed in our study. Thirdly, compression tests of conventional metal foams of low initial density, comparable to our samples, exhibit a plateau in flow stress while the deformation is carried by localized crush bands \cite{SimoneGibson1998,GibsonAshby1999,Michailidis2011b}. NPG lacks the plateau, and in fact an experimental investigation of its microstructure evolution during compression testifies to uniform deformation as opposed to crush bands \cite{Jin2009}.

Our discussion of the distinctive behavior of NPG emphasizes the role of its nanoscale microstructure. As one related aspect, it is well established that the strength of nanoscale objects typically increases with decreasing size \cite{Uchic2004,Volkert2006a,Greer2011} and high strength comes along with low strain hardening \cite{Dou2008}. As a network of nanoscale ligaments, NPG is expected to exhibit a similar trend in response to external load. Yet, in line with previous experimental study \cite{Jin2009}, our results show that both real and virtual NPG samples are highly deformable, with ill-defined yield strength, extremely high compliance and pronounced strain hardening. Here, we address the most relevant issues.

\subsection{Surface-induced stress and initial dislocation content}

The microstructure of our virtual sample was generated by simulating spinodal decomposition on a rigid crystal lattice. By construction, the initial structure was therefore free of strain and of lattice defects. Prior to the MD studies of deformation behavior, the virtual sample was subjected  to a relaxation free of load. The ensuing densification by plastic deformation, accompanied by the creation of various lattice defects, may be understood as the consequence of surface stress. The surface of a solid exerts forces in the underlying bulk (see below) that need to be compensated by bulk stress. For elongated structures, such as the ligaments of NPG, the surface-induced bulk stress has a shear component \cite{Weissmueller1997} that may prompt dislocation nucleation and spontaneous plasticity. Experiments on NPG detect the associated volume shrinkage and the generation of dislocations \cite{Parida2006}, and atomistic simulation has previously confirmed the mechanism \cite{Crowson2009}.

While Fig.~\ref{fig:2} documents the formation of these surface-induced stresses during the energy relaxation of the initially unstrained NPG sample, the magnitude of  the surface-induced stress, $\mathbf S$, in the bulk of the nanoporous material is governed by the generalized capillary equation for solids \cite{Weissmueller1997},
\begin{equation}
\label{eqn:1}
\int \mathbf S \mathrm d V + \int \mathbf s \mathrm d A = 0 \,,
\end{equation}
where $V, A$, and $\mathbf s$ denote volume, surface area and surface stress tensor, respectively. This equation embodies a trend for even isotropic surface stress to create anisotropic stress in bulk, whenever the surface orientation distribution function is anisotropic. Approximating the surface stress as isotropic with magnitude $f$, so that $\mathbf s = f \mathbf P$ with $\mathbf P$ the surface projection tensor \cite{Gurtin1998}, and considering a cylindrical nanowire as a cartoon of the ligament geometry, one obtains the principal values of surface-induced bulk stress in axial and radial directions, respectively, as $S_{\mathrm A} = - 4 f / L $ and $S_{\mathrm R} = 2 f / L $ \cite{Weissmueller1997}. The maximum projected shear stress, of magnitude $\tau = 2 f /L$, is found on planes inclined by $45^{\circ}$ to the wire axis. For the EAM potential used in this study $f\approx 1.1\:\mathrm{J/m^2}$ \cite{Swaminarayan1994}. With a ligament diameter of $L = 3.15\:\mathrm{nm}$, the surface-induced shear stress is thus estimated at $\tau \approx 0.70\:\mathrm{GPa}$. While this value is well below the theoretical shear strength of our potential, the disordered nature of the nanoporous network entails variations in ligament diameter and a fraction of local cross-sections that are well below $3.15$~nm in diameter and that, therefore, may experience local surface-induced stresses of sufficient magnitude to induce spontaneous plastic shear.

\subsection{Early stages of plasticity under load}

The surface-induced plastic deformation during the equilibration has important ramifications for the yield strength of the material. Most obviously, at room temperature, the virtual samples start out with a significant initial dislocation density. It has been well documented that the mechanical behavior of nanoscale objects depends crucially on this parameter. While practically ideal strength may be achieved in dislocation-free whiskers, irrespective of their size, nanoscale pillars with an initial dislocation content are found considerably weaker \cite{Bei2008,Lee2009,Jennings2010}. The immediate yield onset of NPG may thus be promoted by its high initial dislocation density. However, the very same behavior observed at nearly zero temperature where {\it no} dislocation was stored in the virtual samples indicates that the pre-existing dislocations are not the only reason for the immediate yielding.

Equally relevant is the observation that not all ligaments  yield under the action of the surface stress during the initial equilibration. This is naturally understood as the consequence of a distribution of resolved surface-induced shear stresses. Only some ligaments will yield, thereby thicken and so reduce  their local surface-induced prestress. Other ligaments may retain their prestress. It is then natural to assume that some ligaments are prestressed close to yielding, with only small external load required to locally reach the theoretical shear stress. In other words, the heterogeneous nature of the surface-induced prestress contributes to early yielding and, when considering the entire porous network, to an extended elastic-plastic transition as the external load increases.

The above conjecture is perfectly born out by the observations on the mechanical behavior of our virtual sample and by the experimental data in Fig.~\ref{fig:3}. The lasting plastic deformation after unload even from strain as small as $\sim 1\%$ testifies to the quasi instantaneous onset of yielding as load is applied, in agreement  with the presence of very large surface-induced prestress in some ligaments. The same picture may also explain the initial increase in flow stress, which may reflect the exhaustion of local configurations with a low yield stress. In other words, as the deformation proceeds, the external load needs to increase in order to overcome those configurations in which the prestress is less active in assisting the deformation. Similar behavior has been discussed in massive nanocrystalline metal when deformation events are confined to individual grains with a distribution of  crystal orientation and hence a distribution of local Schmid factor \cite{Li2009}. Stress-strain curves of such nanocrystalline metal samples exhibit an extended elastic plastic transition, yet with a preceding, well expressed elastic regime. The complete lack of a regime of purely elastic deformation is characteristic of nanoporous gold.

The disordered nature of the porous network is another reason for stress heterogeneity in nanoporous gold. Depending on its orientation relative to the macroscopic load, and depending on the load-bearing capacity of its neighboring ligaments, each ligament will suffer compressive load or bending moments which are related to the macroscopic load by a local stress concentration factor. The wide distribution of stress concentration factors contributes to the gradual yield onset. This distribution should be well approximated by the FEM simulation of NPG in Ref. \cite{Huber2014}. Yet, the simulated stress strain curves of that study do not reproduce the immediate yielding of the present work. The existence of an extended regime of purely elastic deformation in the FEM simulation thus argues against  distributed stress concentration factors as the origin of early yielding. These considerations again emphasize the importance of the surface-induced prestress that was discussed above.

As another point of support for the above argument we advertise its implication of a  compression-tension asymmetry. The surface-induced bulk stress would enhance compressive load but counteract tensile load. The strength should thus be enhanced in tension. This was pointed out by Farkas {\it et al.} \cite{Farkas2013}, who indeed confirmed the asymmetry in their MD simulation. In this relation we point out that a well-expressed linear elastic regime is seen in MD tension tests of nanoporous gold, but when compression is simulated with the same potential the elastic regime is largely diminished \cite{Farkas2013}. This is again compatible with surface-induced prestress as the origin of yielding in compression but not in tension.

Surface roughness is another factor that might contribute to the deformability of NPG. For nanowires and nanopillars, MD simulations suggest that step edges and corners result in stress concentration that promotes dislocation nucleation and thus reduces the yield strength \cite{Brochard2000,Diao2004,Hyde2005}. Recently, the local strain concentration at edges of nanostructures could be experimentally confirmed by imaging with aberration-corrected TEM \cite{Roy2014}. In view of the impact of roughness for yielding it is significant that Fig.~\ref{fig:1}b reveals an abundance of step edges and corners in NPG; indeed at least part of these defects are geometrically necessary in order to achieve the nanoscale surface curvature of  the ligament network.

\subsection{Strain hardening}

The comparison between the evolution of flow stress with strain in experiment and simulation (Fig.~\ref{fig:3}) reveals a remarkable agreement between both data sets. This suggests that the simulation catches the essentials of the strain hardening behavior of nanoporous gold.

The strain hardening of NPG may partly be understood as a manifestation of densification under compression. As pointed out in Section 3.3, the lack of transverse plastic strain implies that compressive axial strain entails a numerically equal volume compression and, hence, an enhanced solid volume fraction. In view of the Gibson-Ashby scaling equation for the variation of flow stress with initial density \cite{Gibson1982},
\begin{equation}
\label{Eq:Scaling_Flowstress}
\sigma ^{\mathrm{flow}} \propto \varphi^{3/2} \,,
\end{equation}
it is natural to also expect a qualitatively similar behavior when the material is densified during compression. In a constitutive sense, the increase in flow stress with ongoing strain represents an (effective, macroscopic) strain hardening. Of course, the link to the scaling law Eq.~(\ref{Eq:Scaling_Flowstress}) can only be qualitative in view of the dissimilarities in structures emanating from changing density during synthesis or during compression, respectively. In this qualitative sense, the expectation is  supported by the observation that suppressing the volume change by infiltrating the pore space of NPG with a polymer will indeed eliminate the strain hardening  \cite{Wang2013}. Yet, the finite element simulation in Ref.~\cite{Huber2014} demonstrates the need to account additionally for work hardening in the constitutive equation  of each nanoscale ligament, on top of the mere densification effect.

Experiment points towards dislocation storage and dislocation interaction as the origin of the local work hardening within each ligament. Electron backscatter diffraction data of the microstructure evolution during deformation of nanoporous gold testify to the gradual evolution of mosaic spread at a scale larger than the ligament size  \cite{Jin2009}. This requires dislocation storage, and the very large increase in dislocation density in our simulation clearly supports this conjecture. What is more, the experimental data reveal a gradual increase of the strain rate sensitivity as the deformation proceeds, an observation that is consistent with an increasing role of  dislocation interaction \cite{Jin2009}. Our simulation supports that picture as well. The final dislocation density of several dislocations per ligament is extremely high, suggesting drastic Taylor hardening. The Lomer-Cottrell locks that are apparent in our deformed structure (as exemplified in Fig.~\ref{fig:8}) provide a direct evidence of dislocation interaction. We note that the continuous accumulation of lattice defects during compression and the impact of dislocation interaction implies that the dislocation-starvation scenario, which is often considered in models of small-scale plasticity \cite{Greer2005,Greer2006}, does not apply for NPG.

The strain rate of MD simulation is much higher compared to that of our experiment. When there are competing deformation mechanisms with different strain-rate sensitivity, as in nanocrystalline metals \cite{Yamakov2004,Meyers2006}, the dominant deformation mechanism depends on the strain rate. Specifically, classical MD simulations may overestimate the contributions of dislocation slip \cite{Schafer2013}. In that respect, we emphasize that our simulation results as well as previous studies \cite{Jin2009,Sun2013,Farkas2013} consistently highlight dislocation slip as the only governing process in deformation of NPG. We therefore expect this mechanism to dominate independently of the strain rate, supporting a discussion of simulation and experiment within the same conceptual framework.

\subsection{Initial compliance and subsequent stiffening}

As is presented in Fig.~\ref{fig:6}, the evolution of $Y^{\mathrm{eff}}$ is characterized by two features: exceptionally high initial compliance and a subsequent stiffening under compression. The excellent consistency between simulation and experiment indicates that the unusual behavior is indeed intrinsic for NPG.

As with the early yielding, the elastic softening must partly be related to the microstructural heterogeneity. Due to the topological disorder, different regions will exhibit different elastic response, and soft regions might contribute to high effective compliance. FEM modeling \cite{Huber2014} confirms the  strong correlation between structural disorder and compliance in NPG. Yet, the FEM model fails to reproduce the strong stiffening during the initial compression. This points towards the role of atomic-scale processes which are not reflected in the constitutive laws of the FEM.

A conceivable atomistic origin of the anomalous compliance is reversible microplasticity based on the bow-out of pre-existing dislocation segments between their anchor points. For an example of this effect, here in cold-worked iron, see \cite{Benito2005}. Yet, the equally high compliance of the essentially defect-free virtual NPG of our low-temperature runs immediately rules out a decisive impact of preexisting dislocations on the compliant response of NPG.

\begin{figure}[h!]
\centering
\includegraphics[scale=1]{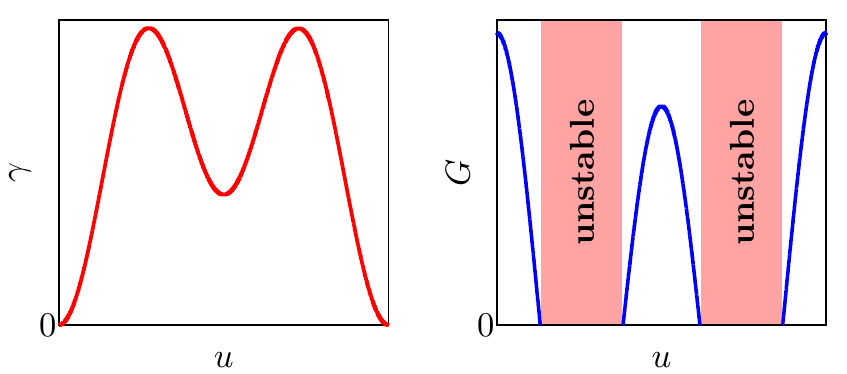}
\caption{\label{fig:9}(Color online) Schematic illustration of shear instability. Graph at left shows generalized stacking fault energy function, $\gamma$, versus shear displacement, $u$, between adjacent atomic planes. Graph at right shows the associated variation of the shear modulus, $G$, which scales with the second derivative of $\gamma(u)$. The negative-valued second derivative between the points of inflection of $\gamma(u)$ implies instability to shear.}
\end{figure}

Similarly to the elasticity of nanowires, a more consistent explanation of the anomalous compliance considers surface-related phenomena. Yet, the origin of size-dependent effective stiffness values of nanowires is controversial, see e.g. Ref. \cite{McDowell2008} and references therein. Surface excess elasticity is an obvious explanation  \cite{Miller2000,Wang2007}, and is supported by the observation of large changes in the stiffness when the surface state is reversibly varied \cite{Mameka2014}, yet an agreement on the sign of the surface excess elastic constants, stiffer or more compliant than bulk, has yet to emerge. The role of elastic nonlinearity of the core of wires prestressed by the capillary forces has also been advertised (e.g. see \cite{Liang2005,Chen2012,Yun2009}); the issue is challenging since the nonlinear elastic response of cubic crystals  is complex, see e.g. \cite{Wang2009}. Irrespective of their details, the above considerations require the modified stiffness to depend on the specific surface area. It is therefore remarkable that our findings show substantial stiffening of NPG after small plastic strain, while the net surface area is essentially invariant. This points towards processes other than surface excess elasticity or modified elastic response of the bulk. One issue stands out in the present geometry: as discussed in Section 4.1, the surface stress induces shear stresses which exceed the theoretical shear strength in some ligaments and closely approach it in others. Since the theoretical shear strength is determined by a point of inflection in the generalized stacking fault energy curve, it entails a vanishing shear modulus. This is illustrated schematically in Fig.~\ref{fig:9}. In other words, the surface-induced stress and the concomitant shear will systematically reduce the shear stiffness of the ligaments in NPG. This is well consistent with the anomalous compliance of the material in its initial state. Furthermore one expects that, as the material is plastically compressed, more and more of these near-unstable configurations will be pushed over to stable faulted or twinned configurations, relaxing their stress and reverting back to a configuration of conventional, high shear stiffness. This agrees with the observed stiffening during the early stages of plastic deformation. Even though the shear instability in this picture seems unusual, it is well supported by all observations of our work.

\subsection{Size-dependent flow stress}

The discussion so far advertises a truly remarkable agreement  between simulation and experiment, including the stiffness and its evolution during compression, the initial yield behavior and the strain hardening. This agreement supports  the notion that the model structure is compatible with experiment in respect to its topological features and likely also in respect to the state of the surface. Yet, the agreement breaks down when it comes to the structure size, where the ligaments of the experimental sample are around tenfold larger than those of the simulation. As a consequence, and  in view of the well-known trend for strength enhancement at reduced size, the numerical agreement between the flow stress values in simulation and experiment is surprising. As the agreement between the stiffness values suggests that the topology of the simulation is realistic, we suspect that the unexpectedly low flow stress  of the simulation is linked to the surface geometry. Specifically, Fig.~\ref{fig:1}b illustrates that the small structure size in the simulation entails extremely high local surface curvatures and, in the consequence, extreme densities of geometrically necessary step edges and kinks. We have already mentioned in Section 4.2 above that atomistic simulation studies \cite{Brochard2000,Diao2004,Hyde2005} point towards a strong influence of  surface defects on the yield behavior. The low flow stress in the simulation supports these remarks, suggesting that surface roughness has an important impact on local stress concentrations and, hence,  on  the macroscopic yield behavior of the extremely small structures of our simulation.

\section{Summary}

We have presented a study of the mechanical behavior of nanoporous gold (NPG) in compression by means of molecular dynamics simulations. A virtual sample prepared by spinodal decomposition with ligament size of $3.15\,\mathrm{nm}$ and solid fraction $\varphi = 0.297$ was uni-axially loaded and thus strained to very large deformation. Experimental data are shown for comparison.

Our study finds close agreement between experiment and simulation, advertising  the following common phenomenology:

\begin{itemize}

\item NPG can be deformed to large strain in compression, with very high strain-hardening coefficient.

\item NPG yields immediately even at the smallest applied load.

\item The plastic Poisson ratio remains essentially zero throughout a large range of strain.

\item As the most remarkable observation, NPG is anomalously compliant, with a stiffness much lower than predicted by the relevant Gibson-Ashby relation.

\item There is a pronounced stiffening in the initial stages of compression.

\end{itemize}

It is also remarkable that both, the effective macroscopic stiffness and the flow stress, agree closely between experiment and simulation, even though the ligament  size of the experimental sample is tenfold larger than that of the simulation. This difference highlights that, in spite of several remarkable similarities between the deformation phenomena in simulation and experiment, it would be inappropriate to over-emphasize the match between the results from the two approaches. Interatomic potentials in MD simulation do not simultaneously provide precise agreement with the surface tensions, surface stresses, and stacking fault energies of the experiment, nor do the accessible timescales of MD allow for a reproduction of the orders of magnitude slower processes in the experiment. Nonetheless, discussing the simulation results on their own in a self-consistent way affords insights into the microscopic processes and their interactions by which nanoporous solids with microstructure similar -- though not identical -- to the real world deform. Based on a critical discussion of these processes in relation to the experiment, our study arrives at the following suggestions with respect to the acting deformation mechanisms:

\begin{itemize}

\item Even in the absence of external load there is local yielding driven by capillary forces. As-prepared samples therefore already exhibit lattice defects.

\item We propose that the surface-induced stresses create a distribution of shear stress that, in its tail, reaches the theoretical shear strength. The regions of somewhat lesser shear stress are close to an elastic instability, and are distinguished by vanishing shear stiffness. This is a key contribution to the anomalous compliance of NPG.

\item Dislocation activity occurs immediately upon loading, and compressive plastic deformation leads to the continuous accumulation of lattice dislocations.

\item The storage of lattice defects and our observations of locks imply that, in addition to densification, standard dislocation-based mechanisms must contribute to the strengthening of NPG under compression. The dislocation-starvation scenario, which is often considered in models for nano-object deformation, does not apply to NPG.

\item Similarly, the evolution of effective Young's modulus during compression is not a simple density-related effect. Other factors, including the relaxation of highly stressed (and, therefore, nearly shear-unstable) regions by faulting, are important.

\end{itemize}

\section*{Acknowledgments}
The authors gratefully acknowledge the Gauss Centre for Supercomputing (GCS) for providing computing time through the John von Neumann Institute for Computing on the GCS share of the supercomputer JUQUEEN at J{\"u}lich Supercomputing Centre. Computing time was also made available by Lichtenberg-High Performance Computer Cluster at TU Darmstadt. JW acknowledges support by SFB 986 ``Taylor-Made Multiscale Materials Systems - $\rm M^3$'', subproject B2.

\section*{Appendix A. Supplementary data}
Supplementary data associated with this article can be found in the online version. For the sake of clarity, the animation Movie-3 only shows reconstructed surface, planar defects (red), and dislocations (green).

\section*{References}
\bibliography{reference}
\bibliographystyle{model3-num-names}
\end{document}